\documentclass[twocolumn,times,tighten]{aastex63}
\usepackage{color}
\usepackage{enumitem}
\usepackage{hyperref}
\usepackage{verbatim}
\usepackage{amsmath,amstext,mathrsfs}
\usepackage[all]{hypcap} 
\usepackage{afterpage}
\usepackage{color}
\usepackage{soul}
\newcommand*{\re}{} % New material

\newcommand*{\torefereeone}{}
\newcommand*{\torefereetwo}{}
\accepted{by ApJ on May 13 2020}	
\begin{document}
\shorttitle{Curvature of B-field lines}
\title{Curvature of magnetic field lines in compressible magnetized turbulence: \\Statistics, magnetization predictions, gradient curvature, modes and self-gravitating media}
%\correspondingauthor{Ka Ho Yuen}

\author[0000-0003-1683-9153]{Ka Ho Yuen}
\affiliation{Department of Astronomy, University of Wisconsin-Madison, USA} 
\email{kyuen@astro.wisc.edu}
\author[0000-0002-7336-6674]{Alex Lazarian}
\affiliation{Department of Astronomy, University of Wisconsin-Madison, USA} 
\affiliation{Korea Astronmy and Space Science Institute, Daejeon 34055, Republic of Korea}
\email{alazarian@facstaff.wisc.edu}

\begin{abstract}
Magnetic field lines in interstellar media have a rich morphology, which could be characterized by geometrical parameters such as curvature and torsion. In this paper, we explore the statistical properties of magnetic field line curvature $\kappa$ in compressible magnetized turbulence. We see that both the mean and standard deviation of magnetic field line curvature obey power-law relations to the magnetization. Moreover, the power-law tail of the curvature probability distribution function is also proportional to the Alfvenic Mach number. We also explore whether the curvature method could be used in the field-tracing Velocity Gradient Technique. In particular, we observe that there is a relation between the mean and standard deviation of the curvature probed by velocity gradients to $M_A$. Finally we discuss how curvature is contributed by different MHD modes in interstellar turbulence, and suggests that the eigenvectors of MHD modes could be possibly represented by the natural Fernet-Serrat frame of the magnetic field lines. We discuss possible theoretical and observational applications of the curvature technique, including the extended understanding on a special length scale that characterize the importance of magnetic field curvature in driving MHD turbulence, and how it could be potentially used to study self-gravitating system.
\end{abstract}

\section{Introduction}

Turbulence is ubiquitous in astrophysical environment and the interstellar gases are permeated by turbulent magnetic fields. Magneto-hydrodynamic (MHD) turbulence plays a very important role in various astrophysical phenomena (see \citealt{Armstrong1995ElectronMedium, Chepurnov2010ExtendingData,2003matu.book.....B,2004ARA&A..42..211E,MO07,2019tuma.book.....B}), including star formation (see \citealt{MO07,MK04,2016ApJ...824..134F}), propagation and acceleration of cosmic rays (see \citealt{J66,2000PhRvL..85.4656C,2004ApJ...604..671F,YL08,2014ApJ...784...38L,2016ApJ...833..131L,2018ApJ...868...36X}), as well as regulating heat and mass transport between different ISM phases (\citealt{1993MNRAS.262..327G, 2000ApJ...543..227D,LP04,LP06,2001ApJ...561..264D,2006ApJS..165..512K,2006MNRAS.372L..33B,2006ApJ...653L.125P} see \citealt{Draine2009} for the list of the phases).

{\torefereeone

The anisotropy of MHD turbulence is being well studied in a number of important theoretical papers \citep{1981PhFl...24..825M,1983PhRvL..51.1484M,1983JPlPh..29..525S,1984ApJ...285..109H} The studies of the MHD turbulence of the solar wind is presented in e.g. \cite{1995SSRv...73....1T} \& \cite{1995ARA&A..33..283G} (see \citealt{2013LRSP...10....2B} for a review). The attempts in estimating the anisotropy from observations of the magnetosphere and solar wind resulted in the development of the model of MHD turbulence (see \citealt{1992JGR....9717189Z} and reference therein) that incorporates the concept of 2D "reduced MHD" perturbations consisting of 2D ”reduced MHD” perturbations carrying approximately 80\% of energy and the ”slab” Alfvenic waves carrying the remaining 20\% of energy (see \citealt{2002PhPl....9.2440M} and references therein). 

The theoretical description of incompressible MHD turbulence that corresponds to numerical simulations was achieved through understanding of both "critical balance" that governs the turbulent motions in the strong turbulence regime \citealt{GS95}, (henceforth GS95)\footnote{\torefereetwo In GS95 the insight by \citep{1984ApJ...285..109H} in terms of magnetized interstellar turbulence was called "nothing short of prophetic". In addition, GS95 study acknowledges that that the "critical balance" between parallel and perpendicular timescales is the key assumption in the derivation of the Straus (1976) equations that was claimed in Montgomery (1982) to describe the anisotropic state of incompressible MHD turbulence. Nevertheless, unlike GS95, the aforementioned papers did not make the final step and did not provided the derivation of the spectra and the parallel and perpendicular scales.}  and the role of  turbulent reconnection that is a part and parcel of the turbulent cascade in \cite{LV99} (henceforth LV99). GS95 predicted that most of the Alfvenic energy is concentrated in the modes with critical balance between the parallel and perpendicular motions leading to the scale dependent anisotropy of the turbulent motions. This anisotropy was derived in GS95 in the mean magnetic field of reference and formulated in terms of scaling relation for the wavenumbers $k_\parallel \propto k_\perp^{2/3}$ with $k_{\parallel,\perp}$ being the parallel and perpendicular wavenumbers respectively. Later research corrected this point by introducing the {\it local system of reference} in which the scale-dependent anisotropy is present. The concept of the local reference system is self-evident from the point of view of turbulent reconnection (see Lazarian et al. 2020 for a review).  It was shown in LV99 that the magnetic reconnection happens over one eddy turnover time and therefore Alfvenic turbulence can be presented as the collection of eddies with their angular velocities aligned with the magnetic field. Naturally, this field is not the mean magnetic field, but the field that surrounds the eddy, i.e. the local magnetic field.\footnote{Incidentally, the critical balance condition in this formulation is a trivial relation between the period of the eddy turnover $\lambda_\bot/v_l$ and the period of the Alfven wave that the rotation of the eddy induces, i.e. $\lambda_\|/V_A$.} Therefore, the turbulence scaling should be studied in respect to the local system of reference. 

The practical way of defining the local system of reference in numerical simulations was suggested first in \citep{CV00} and this and subsequent numerical studies unambiguously confirmed that the critical balance exist only for the eddies, which parallel and perpendicular scales are measured in respect to the local magnetic field and not in respect to the mean magnetic field \citep{CV00,MG01,2002ApJ...564..291C}. To reflect this in formulating of MHD theory, the anisotropy is given as relation between the sizes of parallel and perpendicular eddies given by $\lambda_\|\sim \lambda_\bot^{2/3}$, which substitutes the relation between the wave vectors $k$ in the original formulation of the critical balance.\footnote{We note that in the frame of the mean magnetic field the anisotropy is different, i.e. $k_\|\sim C k_{\bot}$, where $C$ is a constant, i.e. there is no anisotropy that changes with the scale (see Cho et al. 2002). Due to historic reasons, due to the original formulation in the pioneering GS95 study, this fact sometimes causes confusion with the researchers searching the scale-dependent anisotropy measuring parallel and perpendicular direction in respect to the mean magnetic field.} The scales $\lambda_\bot$ and $\lambda_\|$ are different from reciprocals of $k_\bot$ and $k_\|$ as they are measured in different systems of reference.

Further studies allowed to extend the original incompressible MHD theory to compressible media  \citep{2002ApJ...564..291C,CL02,CL03,2010ApJ...720..742K}. A detailed discussion of the theory of turbulence with derivations of the scaling relations as well as the discussion of the stages of the theory development can be found in a reviews (Brandenburg \& Lazarian 2013, Beresnyak \& Lazarian ...) as well in a recent monograph on MHD turbulence (Beresnyak \& Lazarian 2019).}  

The geometry of magnetic field is also a very important way in characterizing its importance in interstellar turbulent media. The Cauchy momentum equation carries a force term that is proportional to ${\bf B}\times (\nabla \times {\bf B})$, where ${\bf B}$ is the magnetic field, which could be decomposed into the pressure term $\nabla({\bf B}^2)$ and the {\it tension term} ${\bf B}\cdot \nabla {\bf B}$ (See \citealt{2003matu.book.....B}). The former term drives compression of fluid elements while the latter term contains information on how magnetic field bending would introduce acceleration to fluid elements. If the magnetic field lines, with their strength being constant, are bent with a curvature $\kappa$, then the tension term would be proportional to $\kappa {\bf B}^{2}$. Therefore characterizing the magnetic field curvature in MHD turbulence allows one to directly estimate how much forces the magnetic field bending exert to fluid elements.

Curvature of magnetic field could also estimate the magnetization. The curvature method has a significant advantage compared to the traditional magnetic field polarization dispersion method which the latter only gives an estimation of $M_A$ in a statistical area. For interstellar media that have force balances within, the curvature of magnetic field is expected to have a proportionality to magnetic field of $B^{-2}$. which reflects the force balance that is employed by by the technique in \cite{2015Natur.520..518L}  that yields results consistent with the \cite{CF53} technique. Notice that the curvature of magnetic field is a local quantity while the method of dispersion could only be measured statistically within a selected region, which could provide significant advantage in characterizing the field strength with higher resolution data. 

This paper {\torefereeone investigates the curvature of magnetic field lines in the case of balanced turbulence.} We start by introducing the theoretical formulation of curvature of a magnetic field line in MHD turbulence, and also the expectation of curvature dependencies on magnetic field strength in terms of MHD turbulence theory. In \S \ref{subsec:math} we discuss the numerical method and introduce an efficient curvature calculation algorithm applicable for both numerically and observationally. \S \ref{sec:stats} we discuss about the statistics on magnetic field line curvature, especially on how it is related to both magnetization and magnetic field strength, both 3D and 2D. In \S \ref{sec:vgt} we discuss the potential use of the curvature of velocity gradients in estimating the field strength. In \S \ref{subsec:modes} we discuss about how the three MHD modes behave both theoretically and numerically when there is a non-zero curvature in magnetic field lines, and the discuss the length scales that curvature would drive the MHD modes. In \S \ref{sec:discussions} we discuss the potential use and caveats of the curvature method. In particular, we discuss how possibly curvature would deduce gravitational status in \S \ref{subsec:gravity}. In \S \ref{sec:conclusion} we conclude our paper.

\section{The theory of magnetic field line curvature}

\label{subsec:math}
\subsection{Mathematical formulation of curvature and torsion of magnetic field lines}
\label{sec:theory}

Magnetic field line can be considered to be "the path" of an imaginary particle with the particle speed ${\bf B}$ at each point in a small neighborhood. Such characterization of "magnetic field lines" globally does not exists since the concept of magnetic field lines becomes ambiguous when we face regions with magnetic reconnections or magnetic field crossing \citep{1958AnPhy...3..347N}. The consideration of the geometry of the local field lines provides immediate advantage: A natural curvilinear frame called {\it Frenet–Serret frame} could be defined locally with two geometric properties about magnetic field called curvature $\kappa$ and torsion $\tau$, which characterize how does the magnetic field lines deviated from a line and a plane respectively. While their velocity counterpart (velocity curvature, torsion) are well studied (See, e.g. \citealt{2006JTurb...7...62B,2011PhRvE..83c6314K}), the study of magnetic field curvature has not been popular until recently \citep{2019PhPl...26g2306Y}.  

In the studies of magnetic field topology, the concept of curvature is important since magnetic field lines are expected to have a smaller curvature as the strength of the magnetic field increases.  A recent numerical work by \cite{2019PhPl...26g2306Y} shows the probability density function (PDF) of the magnetic field curvature has a power-law tail of $\kappa^{-2}$ in 2D and $\kappa^{-2.5}$ in 3D for incompressible simulation with initially fluctuation energy equi-partitioned between the kinetic and magnetic ones. They also show magnetic field lines follow a proportionality relation of $\kappa \propto f_B/B^2$ with $f_b$ represents the normal force component. Observationally \cite{2015Natur.520..518L} uses the curvature of magnetic field lines as an estimate of magnetic field strength in NGC6334. Together, curvature of magnetic field lines becomes an important physical quantity in characterizing the strength of magnetic field.

Mathematically, we would parametrize the magnetic field lines by the line variable $s$ assuming the magnetic field forms a vector field for a imaginary particle. Intuitively, we would imagine a particle placed at at an arbitrary initial position $\bf{r}_0$ and allow it to evolve following the vector integral $\frac{d{\bf L}_B}{dt} = {\bf B}({\bf{r}-\bf{r}_0})$ with the magnetic field ${\bf B}$ is a velocity field of the particle. The path length of the imaginary particle is then:
\begin{equation}
    s(l) = \int_0^l dl' \sqrt{B_x^2(l')+B_y^2(l')+B_z^2(l')}
\end{equation}
the Frenet-Serret frame of the the magnetic fields lines would be (Callen 2003):
\begin{equation}
    \begin{aligned}
    \frac{d\hat{t}}{ds} &= &+\kappa \hat{n}&\\
    \frac{d\hat{n}}{ds} &= -\kappa \hat{t}&&+\tau \hat{b}\\
    \frac{d\hat{b}}{ds} &= &-\tau \hat{n}&\\
    \end{aligned}
    \label{eq:FSF}
\end{equation}
where $\hat{t} = \hat{B}$ the unit vector of the magnetic field, $\hat{n},\hat{b}$ are the normal and the binormal vector respectively, and the line derivative $\frac{dx}{ds} = \hat{t}\cdot\nabla x$ if a vector $x$ is a unit vector, $\tau$ is the torsion of the magnetic field line (\S \ref{subsec:torsion}). Under this formulation, the signed curvature could be given by:
\begin{equation}
    \begin{aligned}
    \kappa &= \hat{n}\cdot\frac{d\hat{t}}{ds} 
    \end{aligned}
    \label{eq:CT}
\end{equation}

\subsection{Prediction of dependencies of curvature to magnetic field strength, sonic and Alfvenic Mach number in MHD turbulence}
\label{subsec:pred}

The Frenet-Serret frame is a natural frame in studying the local geometry of magnetic field. We expect that in the presence of magnetized turbulence, there should be a power law $\kappa \propto B^{-\gamma}$ where $\gamma$ is a constant yet to be defined.  From Eq.\ref{eq:CT}, apparently the magnetic field line {\torefereeone curvature} is inversely proportional to the squared amplitude of magnetic field. Indeed, as argued in \cite{2019PhPl...26g2306Y}, if the force term is constant, then $\kappa \propto B^{-2}$ but in the case of MHD turbulence the interaction of velocity and magnetic fields would tend to reduce the dependencies of $\kappa$ to $B$. Therefore we would expect $\gamma<-2$ the properties of turbulence is characterized not only by magnetic field strength but also the velocities and densities of the fluid elements, a more appropriate estimate would be 
\begin{equation}
    \kappa \propto M_A^{\gamma},
    \label{eq:prediction}
\end{equation}
where $M_A=v_{inj}/v_A$ is the ratio of the injection and Alfven velocities, $v_A=B_0/\sqrt{4\pi\rho_0}$.

We can make similar estimation on the dependence of $\kappa \propto M_A^\gamma$ from the theory of MHD turbulence \citep{GS95,LV99}, henceforth GS95 and LV99, respectively. As the \cite{GS95} formulated for trans-{\torefereeone Alfvenic}, i.e. $M_A=1$ turbulence, in what follows we are mostly using the \cite{LV99} expressions obtained for $M_A<1$. For $M_A>1$ the GS95 approach can be easily generalized (see Lazarian 2006) as we discuss this below. 

In the incompressible limit, the parallel and perpendicular length scales of the turbulent eddies would be related by the following relations based on the theory of MHD turbulence. In the following we shall only consider cases where we have either strong magnetic field ($M_A = v_{inj}/v_A<1$, $v_{inj}$ is the injection velocity and $v_A$ is the Alfven velocity) or we are in the regime of dynamically important magnetic field ($M_A>1$ but the lengths scale $l<l_A=L_{inj}M_A^{-3}$, where $L_{inj}$ is the injection scale). Here we consider that the magnetic field eddies are coherent to that of the velocity eddies, so that one could simply use the same scaling law for magnetic field structures without further approximations.

In the cases of strong magnetization, there exists a scale $l=L_{inj} M_A^2$ such that scales smaller than that would be the strong turbulence. We shall spare the discussion of length scales larger than that since that usually have a limited spatial range. We stress that the parallel $\lambda_{\parallel}$ and perpendicular length scale $\lambda_{\perp}$ are defined in terms of the {\it local} direction of magnetic field. {\torefereetwo While in earlier works \citep{1983JPlPh..29..525S,1984ApJ...285..109H,1996JGR...101.7619M} the anisotropy is observed and theoretically tested, these anisotropies are generally measured along the mean magnetic field. In fact, the concept of the local magnetic field is absent in \cite{GS95} where all the closure relations used for the derivation are formulated in the reference frame of the {\it mean} field \citep{LV99}. { The concept of local reference frame was shown to be valid numerically (see \citealt{CV00,MG01,Cho2001SimulationsMedium}, or the appendix of \citealt{LYH18} for a comprehensive discussion.)}. It is important to stress that the relations between  $\lambda_{\parallel}$ and  $\lambda_{\perp}$ are {\it not valid} if $\|$ and $\bot$ distances are measured in respect to the mean field \citep{CV00}.} 

{\torefereetwo   Properties of MHD turbulence are easy to understand within the model of magnetic eddies aligned with the local direction of magnetic field  \citep{LV99} is essential. That means that the anisotropy should be computed in respect to magnetic field at the scale of the eddies. In the local system of magnetic field of the eddies, the parallel and perpendicular scales of the eddy are related for $M_A<1$ as} \citep{LV99}:
\begin{equation}
\lambda_{\parallel} \sim L_{inj} \left(\frac{\lambda_{\perp}}{L_{inj}}\right)^{2/3} M_A^{-4/3}, 
\label{eq:scaling_smallma}
\end{equation}
{\torefereetwo where $\lambda_{\parallel,\perp}$ are the parallel and perpendicular length scales of the turbulent eddies under \cite{LV99} formalism. Eq. (\ref{eq:scaling_smallma}) has two differences from the original expression by \cite{GS95}. First of all, it relates the physical scales of the eddies rather than wavenumbers. The latter are given in the frame of the mean field and do not exhibit the scale dependent anisotropy of Eq. (\ref{eq:scaling_smallma}). Second, \citeauthor{GS95} theory is formulated for $M_A=1$.  } 

{\torefereetwo Readers should be careful that the parallel and perpendicular scales that we are discussing here are referring to the local scales argument in \cite{LV99}. Numerically the $2/3$ scaling law is tested in \cite{CV00}, and the energy spectrum is tested in \cite{CL03}. Those studies confirmed that the aspect ratio is larger for smaller eddies and, in fact, follows the predicted “critical balance” relation. The justification of such a procedure follows from the eddy description of MHD turbulence based on \cite{LV99} study. The subsequent studies, see \cite{2002ApJ...564..291C,2005ApJ...624L..93B}, provided the numerical support for these scalings. Recently the appendix of \cite{2018ApJ...865...54Y} also revisit this issue and state again clear that only the local computation would yields the desired 2/3 scaling law, in agreement with the expectations based on \cite{LV99} reconnection theory.}

For super-Alfvenic turbulence $M_A>1$ and at large scales magnetic fields do not change the Kolmogorov picture.While in the case of weak magnetization and $l<l_A=L_{inj}M_A^{-3}$, the parallel $\lambda_{\parallel}$ and perpendicular length scale $\lambda_{\perp}$ are related by:
\begin{equation}
    \lambda_{\parallel} \sim L_{inj} \left(\frac{\lambda_{\perp}}{L_{inj}}\right)^{2/3} 
\label{eq:scaling_largema}
\end{equation}

In either cases, the {\it minimal} curvature of the eddies measured for the eddy of size $\lambda_\perp$ could be estimated by $\lambda_\perp/\lambda_\parallel^{2}$. Therefore we would obtain the expected curvature for eddies {\it as a variable of the minor length $\lambda_{\perp}$}:
\begin{equation}
    \kappa(\lambda_\perp) \sim L_{inj}^{-2/3}\lambda_\perp^{-1/3} \cdot
    \begin{cases}
    M_A^{8/3} & (M_A<1, l<L_{inj}M_A^2)\\
    1         & (M_A>1, l<L_{inj}M_A^{-3})
    \end{cases}
    \label{eq:kappa_ma}
\end{equation}
which the curvature at length scale $\lambda_{\perp}$ is proportional to the Alfvenic Mach number. {\torefereetwo Here the curvature that we measured, $\kappa(\lambda_\perp)$, correspond to the curvature we measure in real space when we only discuss eddies of size $\lambda_\perp$.} For eddies larger than the scales in the two individual cases, i.e. for $l>L_{inj}M_A^2$ if $M_A<1$ and $l<L_{inj}M_A^{-3}$ for $M_A>1$ respectively, the curvature of these {\it isotropic} eddies with scale $\lambda$ is simply $\lambda^{-1}$ with no relation between $\lambda$ and $M_A$. 

The measured curvature value in real space would be {\torefereeone mainly contributed by two factors}: (1) the largest scale that has turbulent anisotropy (2) contributions from eddies with scales larger than the largest scale that has turbulent anisotropy. {\torefereetwo Since the curvature value of real space is basically a statistical sum of all curvature values from the eddies with different size, it is natural to consider for what size the eddy dominates over the measurements. In the case of sub-Alfvenic turbulence, the largest scale measurable with anisotropy is $l=L_{inj}M_A^2$. Therefore the curvature of eddies at scale $l=L_{inj}M_A^2$ would be:}
\begin{equation}
    \kappa(\lambda_\perp=L_{inj}M_A^2) = L_{inj}^{-1} M_A^2 \quad (M_A<1)
    \label{eq:kappa_maleone}
\end{equation}
while for super-Alfvenic turbulence, the respective transition scale is $l=L_{inj}M_A^{-3}$
\begin{equation}
    \kappa(\lambda_\perp=L_{inj}M_A^{-3}) = L_{inj}^{-1} M_A^1 \quad (M_A>1)
    \label{eq:kappa_maleone}
\end{equation}
{\torefereetwo The contribution of curvature from eddies with scales larger than the scale having turbulence anisotropy has no dependence of $M_A$. The observed curvature would then be a statistical sum of curvature values acting on different size of eddies.} Moreover, notice that the discussion above is based on the turbulence scaling laws from incompressible turbulence theory. {\torefereeone Since compressible modes in compressible magnetized turbulence (See \citealt{CL03}) rarely have non-zero curvature and the observed curvature is the statistical average of the curvature values obtained from the three MHD modes, the larger weight of compressible modes would decrease the observed curvature} (See discussion in \S \ref{subsec:modes}). {\torefereeone Observationally we cannot perform an analysis of curvature as a function of $\lambda_\perp$ unless precise magnetic field orientations are given, which requires the knowledge of the knowledge of the 3D magnetic field distribution.  Statistically one would only obtain a value of curvature contributed by all three modes of MHD turbulence, and all scales. } As a result we expect that the measured curvature value has a weaker dependence to $M_A$ for both cases, i.e. we expect :
\begin{equation}
    \kappa \sim L_{inj}^{-1} \cdot
    \begin{cases}
    M_A^{2-\alpha} & (M_A<1)\\
    M_A^{1-\alpha} & (M_A>1)
    \end{cases}
    \label{eq:kappa_ma}
\end{equation}
for some constant $\alpha$ dependent on the properties of injection and the composition of modes. {\torefereeone As we will see in \S \ref{sec:stats}, the inclusion of $\alpha$ is necessary in explaining the behavior of curvature in compressible turbulence.}

We also expect the index $\gamma = 2-\alpha (M_A<1) \text{ or } 1-\alpha (M_A>1)$ to be shallower in 2D, i.e $\gamma_{2D} < \gamma_{3D}$ since the projection effect of magnetic field would tends to cancel out the magnetic field deviations that is not a straight line. As a result the curvature observed in the projected space should be systematically smaller than that in the 3D.

\section{Method}
\label{sec:method}
\subsection{Simulation setup}
The numerical data cubes {\re are} obtained by 3D MHD simulations that is from a single fluid, operator-split, staggered grid MHD Eulerian code ZEUS-MP/3D to set up a three dimensional, uniform turbulent medium. Our simulations are isothermal with $T=10K$. To simulate the part of the interstellar cloud, periodic boundary conditions are applied. We inject turbulence solenoidally\footnote{  These simulations are the Fourier-space forced driving isothermal simulations. {  The choice of force stirring over the other popular choice of decaying turbulence is because only the former will exhibit the full characteristics of turbulence statistics (e.g power law, turbulence anisotropy) extended from $k=2$ to {\re a dissipation scale of $12$ pixels} in a simulation , and matches with what we see in observations (e.g. \citealt{Armstrong1995ElectronMedium, Chepurnov2010ExtendingData}) }.} 

For our controlling simulations parameters, various Alfvenic Mach numbers $M_A=V_{inj}/V_A$ and sonic Mach numbers $M_s=V_{inj}/V_s$ are employed \footnote{ For isothermal MHD simulation without gravity, the simulations are scale-free. The two scale-free parameters $M_A,M_s$ determine all properties of the numerical cubes and the resultant simulation is universal in the inertial range. That means one can easily transform to whatever units as long as the dimensionless parameters $M_A,M_s$ are not changed.}, where $V_{inj}$ is the injection velocity, while $V_A$ and $V_s$ are the Alfven and sonic velocities respectively, which they are listed in Table \ref{tab:sim}. For the case of $M_A<M_s$, it corresponds to the simulations of turbulent plasma with thermal pressure smaller than the magnetic pressure, i.e. plasma with low confinement coefficient $\beta/2=V_s^2/V_A^2<1$. In contrast, the case that is $M_A>M_s$ corresponds to the magnetic pressure dominated plasma with high confinement coefficient $\beta/2>1$. 

From now on we refer to the simulations in Table \ref{tab:sim} by their model name. For example, the figures with model name indicate which data cube was used to plot the corresponding figure. Each simulation name follows the rule that is the name is with respect to the varied $M_s$ \& $M_A$ in ascending order of confinement coefficient $\beta$. The selected ranges of $M_s, M_A, \beta$ are determined by possible scenarios of astrophysical turbulence from very subsonic to supersonic cases.

\subsection{Synthetic observations}
\label{subsec:syn}
The raw data from simulation cubes are converted to synthetic maps for studies of curvature. Since we would investigate both polarization and magnetic field probed by Velocity Gradient Technique (\citealt{YL17a,LY18a}, see \S \ref{subsec:vgtintro}), we shall deliver the method in synthesizing the Stokes parameters, intensities, centroids and velocity channels here.

The Stokes parameters for synthetic observation are given by 
\begin{equation}
\label{eq.stokes}
    \begin{aligned}
    Q&\propto\int dz n \cos(2\theta)\sin^2\gamma_{inc}\\
    U&\propto\int dz n \sin(2\theta)\sin^2\gamma_{inc} \\
    \theta_{pol} &= \frac{1}{2} \text{tan2}^{-1}(U/Q)
    \end{aligned}
\end{equation}
where $n$ is the number density, $\theta,\gamma_{inc}$ are the 3D planer angle and inclination angle of magnetic field vectors with respect to the line of sight respectively. The dispersion of the polarization angle $\theta_{pol}$ is directly proportional to the perpendicular Alfvenic Mach number $M_{A,\bot}$. 

The normalized velocity centroid $C({\bf R})$ in the simplest case\footnote{Higher order centroids are considered in \cite{YL17b} and they have $v^n$, e.g. with $n=2$, in the expression of the centroid. Such centroids may have their own advantages. However, for the sake of simplicity we employ for the rest of the paper $n=1$.} is defined as
\begin{equation}
\begin{aligned}
C({\bf R}) &=I^{-1} \int \rho_v({\bf R},v) v  dv,\\
I({\bf R}) &= \int \rho_v({\bf R},v)  dv,
\end{aligned}
\label{centroid}
\end{equation}
where $\rho_v$ is density of the emitters in the Position-Position-Velocity (PPV) space, $v$ is the velocity component along the line of sight and ${\bf R}$ is the 2D vector in the pictorial plane. The integration is assumed to be over the entire range of $v$.  Naturally, $I({\bf R})$ is the emission intensity. The $C(R)$ is also an integral of the product of velocity and line of sight density, which follows from a simple transformation of variables (see \citealt{2003ApJ...592L..37L}). For constant density, $C(R)$ is just the line of sight velocity averaged over the line of sight. 

We also consider the velocity channel at $v=0$ here as a case study. Mathematically, the density in PPV space of emitters with local sonic speed $c_s({\bf x}) = \sqrt{\gamma k_B T/\mu_{MMW}}$, where $\mu_{MMW}$ is the mean molecular weight of the emitter, moving along the line-of-sight with stochastic turbulent velocity $u(\bf x)$ and regular coherent velocity, e.g. the galactic shear velocity, $v_{\mathrm{g}}(\bf x)$ is \citep{LP04}
\begin{equation}
\rho_s(\mathbf{X},v) =\int_0^S \!\!\! dz 
\frac{\rho(\mathbf{x})}{\sqrt{2\pi \beta_T}}
\exp\left[-\frac{(v-v_{\mathrm{g}}(\mathbf{x}) -
u({\bf x}))^2} {2 c_s^2({\bf X} ,z) }\right] 
\label{eq:rho_PPV}
\end{equation}
where sky position is described by 2D vector $\mathbf{X}=(x,y)$ and $z$ is the line-of-sight coordinate, $\gamma$ is the adiabatic index, $S$ is the cloud depth. Notice that $c_s$ would be a function of distance if the  emitter is not isothermal. The Eq.~(\ref{eq:rho_PPV}) is \textit{exact}, including the case when the temperature of emitters varies in space. The observed velocity channel at velocity position $v_0$ and channel width $\Delta v$ is then, assuming a constant velocity window $W(v)=1$ and $v_g({\bf x})=0$:
\begin{equation}
\begin{aligned}
Ch(\mathbf{X};v_0,\Delta v) 
= &\int_{v_0-\Delta v/2}^{v_0+\Delta v/2}dv \rho_s(\mathbf{X},v)\\
= \int_0^S dz \frac{\rho(\mathbf{x})}{\sqrt{2\pi c_s^2}} &\int_{v_0-\Delta v/2}^{v_0+\Delta v/2} dv e^{-\frac{(v - u({\bf x}))^2} {2 c_s^2 }}
\end{aligned}
\label{eq:channel}
\end{equation}
We shall deliver the methods of computing the gradients of velocity channel and use it as a probe of tracing magnetic fields.

\subsection{Velocity Gradient Technique}
\label{subsec:vgtintro}

The Velocity Gradient Technique (VGT, see \citealt{GL17}) is an innovative method that uses the properties of turbulence anisotropy in MHD turbulence to probe the direction of magnetic field. The basic idea is to obtain the magnetic field predictions by rotating the output the sub-block averaging of the gradients of observables by $90^o$ \citep{YL17a}, {\torefereeone which is supported by a number of theoretical and numerical works \citep{GS95,LV99,MG01,CL02,CL03} } The VGT is applicable to velocity centroids \citep{GL17,YL17a}, intensities \citep{YL17a,IGVHRO} and also velocity channel maps \citep{LY18a}. The same method is also migrated to synchrotron studies and applicable to both synchrotron intensities \citep{Letal17} and multi-frequency synchrotron polarization \citep{2018ApJ...865...59L}.

The gradients of the three observables (intensity, centroid, velocity channel) we introduced in \S \ref{subsec:syn} would be computed as follows. We would first compute the Sobel kernel of the observables which we would call it the raw gradients. The distribution histogram peak of raw gradients would provide us the predicted direction of magnetic field directions probed by the gradients of observables, provided that the Gaussian fitting requirement stated in \cite{YL17a} is satisfied, which is called sub-block averaging in \cite{YL17a}. The $90^o$ rotated gradients are our predicted magnetic field directions by the gradients of observables. We would not apply further improvements of the technique (e.g. \citealt{LY18a,PCA}) since these techniques tend to straighten the estimated magnetic field lines.

\begin{table}
\centering
\begin{tabular}{c c c c c}
Model & $M_S$ & $M_A$ & $\beta=2M_A^2/M_S^2$ & Resolution \\ \hline \hline
huge-0                  & 6.17  & 0.22 & 0.0025 & $792^3$ \\
huge-1                  & 5.65  & 0.42 & 0.011 & $792^3$ \\
huge-2                  & 5.81  & 0.61 & 0.022 & $792^3$ \\
huge-3                  & 5.66  & 0.82 & 0.042 & $792^3$ \\
huge-4                  & 5.62  & 1.01 & 0.065 & $792^3$ \\
huge-5                  & 5.63  & 1.19 & 0.089 & $792^3$ \\
huge-6                  & 5.70  & 1.38 & 0.12 & $792^3$ \\
huge-7                  & 5.56  & 1.55 & 0.16 & $792^3$ \\
huge-8                  & 5.50  & 1.67 & 0.18 & $792^3$ \\
huge-9                  & 5.39  & 1.71 & 0.20 & $792^3$ \\ \hline \hline
\end{tabular}
\caption{\label{tab:sim} Description of MHD simulation cubes {  which some of them have been used in the series of papers about VGT \citep{YL17a,YL17b,LY18a,2018ApJ...865...59L}}.  $M_s$ and $M_A$ are the R.M.S values at each the snapshots are taken. }
\end{table}

\subsection{Self-consistent curvature and torsion obtaining method through enumerating Lagrangian particles}
\label{subsec:lagrangian}

Computing curvature by Eq.\ref{eq:FSF} is difficult since the computation of ${\bf B}\cdot\nabla {\bf B}$ does not often yield a vector parallel to $\hat{n}$ numerically. The reason behind is because the spatial derivative of magnetic field is not guaranteed to have $({\bf B}\cdot\nabla {\bf B}) \cdot {\bf B}=0$. In view of that \cite{2019PhPl...26g2306Y} uses the expression $\kappa = |{\bf \hat{B}}\times ({\bf \hat{B}}\cdot\nabla {\bf \hat{B}})|$ to extract the curvature only part of magnetic field. Below we discuss an algorithm that could possibly bypass the problem with the cost of interpolation accuracy: We treat the magnetic field as the velocity field of an imaginary particle (see \S \ref{subsec:math}) and find the path integral in a sufficiently small area, so that one could have a representation of the "magnetic field line function ${\bf L}_B(t)$" that has:
\begin{equation}
    \begin{aligned}
    {\bf L}_B(t=0) &= {\bf r}_0 =  \text{Spatial position of the pixel}\\
    \frac{d{\bf L}_B}{dt} &= {\bf B}({\bf{r}-\bf{r}_0})
    \end{aligned}
    \label{eq:integrator}
\end{equation}
{\torefereeone Readers should be reminded that if we put $\frac{d{\bf L}_B}{dt} = {\bf v}$, then we are effectively converting the Eulerian hydrodynamic variables to the Lagrangian one, which is simply the  standard Lagrangian particle-tracking algorithm (see \citealt{2006NJPh....8..102O,2007PhRvL..98e0201X}).}

We perform an integrator method that integrates Eq.\ref{eq:integrator} by the Runge–Kutta method (order 2/4/4.5 depending on the accuracy requirement). We first impose an interpolation field so that the magnetic vector field ${\bf B}({\bf r})$ is well defined in a local patch of the 3D real space ${\bf r} \in \Delta L^3 \ \subset \mathbb{R}^3$. The interpolation is usually performed using the family of splines to estimate values between the grid points. Cubic spline is a popular option. The interpolation  could be done very easily through Julia's Interpolation package \footnote{https://github.com/JuliaMath/Interpolations.jl}. Then one could obtain the tangent vector:
\begin{equation}
    {\bf T}(t) = \frac{\dot{\bf L}_B(t)}{|\dot{\bf L}_B(t)|}
\end{equation}
Following Eq.\ref{eq:CT}, one could obtain both $\kappa$ and $\tau$ very easily.

The number of steps required for the integrator method is explicitly depending on how one express the differential operator $\frac{d}{dt}$. For instance, if one uses the one-dimensional Five-point stencil in estimating the differentiation on the position vector $\bf r(t)$ at time $t_0$ with some custom step size $dt$
\begin{equation}
{\bf r}'(t_0) = \frac{-{\bf r}(t_0+2dt)+8{\bf r}(t_0+dt)-8{\bf r}(t_0-dt)+{\bf r}(t_0-2dt)}{12dt}
\end{equation}
then the number of points required in obtaining $\kappa(t=t_0)$ and $\tau(t=t_0)$ (See \S \ref{subsec:torsion}) at position $t_0$ is $2\times (5-1)+1=9$ and $3\times (5-1)+1=13$ points respectively. The advantage of the current method is that we are sticking to the definition of Frenet-Serret frame based on the tangent of the (magnetic field) line. The small $dt$ will guarantee that the integrated line would only be locally defined. The method is valid for both 2D and 3D. However readers should be reminded that torsion is nonzero only when we are tackling a 3D line.

This method is very versatile since one could also compute curvature of scalar structures, say intensity map $I$, through replacing {\bf B} in Eq.\ref{eq:integrator} to the gradients of $I$ rotated by $90^o$. This application is especially useful for the studies based on the Velocity Gradient Technique since curvature is an important quantity aside from orientation and amplitudes of gradients that characterizes the underlying magnetic field properties.

Readers should be reminded that, to compute curvature one must make sure the respective circle that has radius $1/\kappa$ could be represented by the grid resolution. One caveat here is that, it is impossible to obtain curvature values larger than $0.5$ (in units of 1/pixel) since the respective circle with radius smaller than $2$ pixels would not be resolved in the numerical grid.

\section{Statistical properties of magnetic field curvature and torsion in 3D MHD compressible turbulence}
\label{sec:stats}

The very first thing to start with would be to examine the statistical properties of the curvature field. We shall discuss the statistics through the simple tools like mean, standard deviations and histograms both in 3D and projected 2D spaces and both along and perpendicular to the field lines. Previous literature that discuss curvatures are mainly focused on the curvature of velocity fields \citep{2006JTurb...7...62B,2007PhRvL..99s4502O,2008PhFl...20f4104O,2011PhRvE..83c6314K}. The study of statistics on  magnetic field curvature is only recently done by Yang 2019 by an incompressible equi-partitioned magnetized turbulence simulation. As commented in \S \ref{subsec:math}, we expect the statistical parameters would give us a dependence of $\kappa \propto M_A^{\gamma}$ for some {\it positive} values of $\gamma$. For the reader's reference, we are expressing the values of $\kappa$ as the function of numerical pixels since there is an explicit upper bound for $\kappa$ in numerical simulations (See \S \ref{subsec:lagrangian}).

\subsection{Curvature of 3D magnetic field}
\label{subsec:3d}

\begin{figure}[th]
\centering
\includegraphics[width=0.96\linewidth]{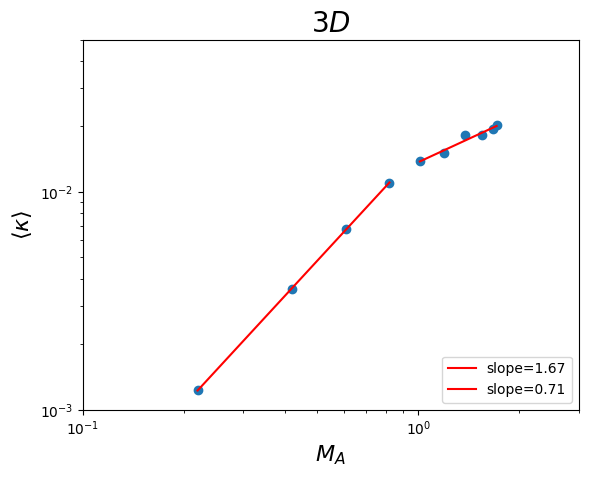}
\includegraphics[width=0.96\linewidth]{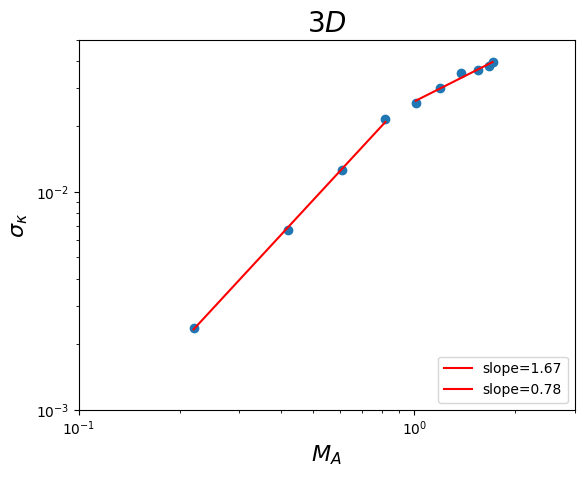}
\includegraphics[width=0.96\linewidth]{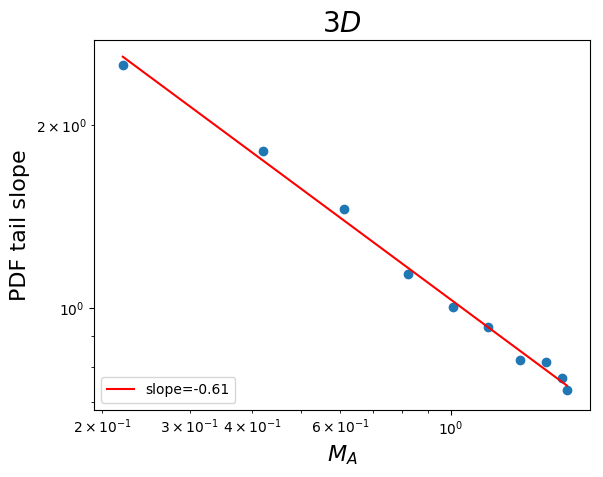}
\caption{\label{fig:curvature_ma_3D} A set of figures showing how the mean value of curvature $\langle \kappa \rangle$ (top), the standard deviation of curvature $\sigma_\kappa$ (middle) and the PDF power-law tail slope of the magnetic field curvature $\kappa$ (bottom) with respect to the Alfvenic Mach number $M_A$. Notice that the power-law tail slope is expected to be negative,  we are plotting its absolute value.  }
\end{figure}

In 3D we do not need to employ the algorithm as listed in \S \ref{subsec:lagrangian} since it is straightforward to show $\kappa = |{\bf \hat{B}} \times ({\bf \hat{B}}\cdot \nabla {\bf \hat{B}})| $. Figure \ref{fig:curvature_ma_3D} shows how the three dimensional statistics would behave as a function of the Alfvenic Mach number. Here we use the mean and standard deviation as ways to extract simple statistics (left and middle of Fig.\ref{fig:curvature_ma_3D}). From our theoretical discussion (\S \ref{subsec:math}), we expect that statistically the magnetic field curvature should be proportional to $M_A^{\gamma}$ for some constant $\gamma$. From the upper plot of Figure \ref{fig:curvature_ma_3D}, we see a two-section power law for $\langle \kappa \rangle$ (in 1/pixel):
\begin{equation}
    \langle \kappa \rangle \propto
    \begin{cases}
     M_A^{1.67} & M_A<1\\
     M_A^{0.71} & M_A\ge 1
    \end{cases}
    \label{eq:meankappafit}
\end{equation}
with a turning point $\langle \kappa \rangle \sim 0.15 \text{pixel}^{-1}$ while for standard deviation of curvature, $\sigma_\kappa$ (middle plot of Fig.\ref{fig:curvature_ma_3D}, in 1/pixel):
\begin{equation}
    \sigma_\kappa \propto
    \begin{cases}
     M_A^{1.67} & M_A<1\\
     M_A^{0.78} & M_A\ge 1
    \end{cases}
    \label{eq:stdkappafit}
\end{equation}
with a turning point $\langle \kappa \rangle \sim 0.25 \text{pixel}^{-1}$. The fitting lines in Eq.\ref{eq:meankappafit} and Eq.\ref{eq:stdkappafit} look surprisingly similar with a relatively high coefficient of determination of 0.9 or above. These measurements are consistent to the theoretical prediction in \S \ref{subsec:pred} with the coefficient $\alpha$ defined in \S \ref{subsec:pred} to be 
\begin{equation}
    \alpha=\begin{cases}
     0.33 & (M_A<1,  \langle \kappa \rangle )\\
     0.29 & (M_A>1,  \langle \kappa \rangle )\\
     0.33 & (M_A<1, \sigma_\kappa)\\
     0.22 & (M_A>1, \sigma_\kappa)
    \end{cases}
\end{equation}
which are fairly similar for these four cases.  The split of the power laws between $M_A\ge1$ and $M_A<1$ are also expected and consistent to our findings in one of our previous work using velocity gradients in probing magnetization \citep{LYH18} since there are different scaling laws below and above $M_A$.

\begin{figure}[th]
\centering
\includegraphics[width=0.96\linewidth]{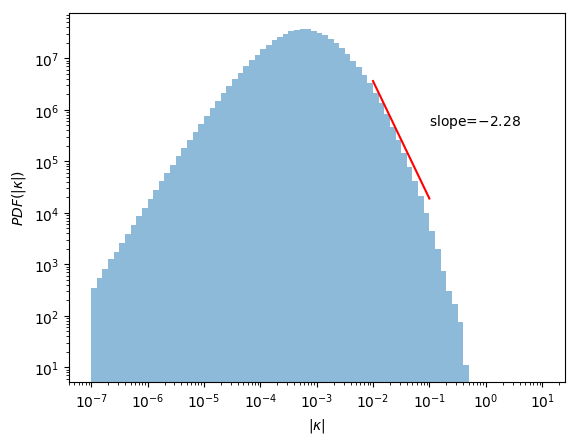}
\caption{\label{fig:pdfillus} A figure showing the PDF of curvature for the cube huge-0. We see a power-law tail slope of $2.28$ for this particular cube, which {\torefereeone has a slightly smaller value than the value recently seen in \citealt{2019PhPl...26g2306Y} (2.5).}}
\end{figure}

We also employ the method of histogram tails since it was suggested that $f(\kappa) \sim \kappa^{-2.5}$ in \cite{2019PhPl...26g2306Y}. The lower part of Figure \ref{fig:curvature_ma_3D} shows the power-law tail slope as a function of $M_A$. One could see that the slope power-law tail is actually a function of $M_A$. Our result should not be directly compared to that in \cite{2019PhPl...26g2306Y} who has a general slope of -2.5 since we are performing simulations for different settings. {\torefereeone (1) we are performing a full 3D, compressible, driven simulations while that of Yang’s paper perform a 3D, incompressible, decaying simulations.  The compressible simulation allows one to store energy into the the two other compressible modes, effectively reducing the curvature of magnetic field lines driven by turbulence driving.  (2)  We are testing our theory using the turbulence statistics, while Yang’s work is based on the formulation of the magnetic force term in the Cauchy momentum equation. (3) We tested the dependencies of the histogram tail as a function of global Alfvenic Mach number with multiple simulations, while\cite{2019PhPl...26g2306Y} is characterizing the curvature as a distribution of local magnetic field strength.} Moreover, it is expected that the PDF of the curvature should be a function of Alfvenic Mach number since in strongly magnetized medium we do not expect a large variation of curvature values due to the constraints of strong field, resulting in a relatively steep power-law tail in the low $M_A$ limit. In the case of weak magnetic field, the allowed values of curvature is less bounded by the magnetic field itself compared to the strong field cases, which would lead to an extended PDF tail and a shallower slope. 

\subsection{2D Magnetic field curvature obtained in synthetic observations}
\label{subsec:stokes}

\begin{figure*}[th]
\centering
\includegraphics[width=0.48\linewidth]{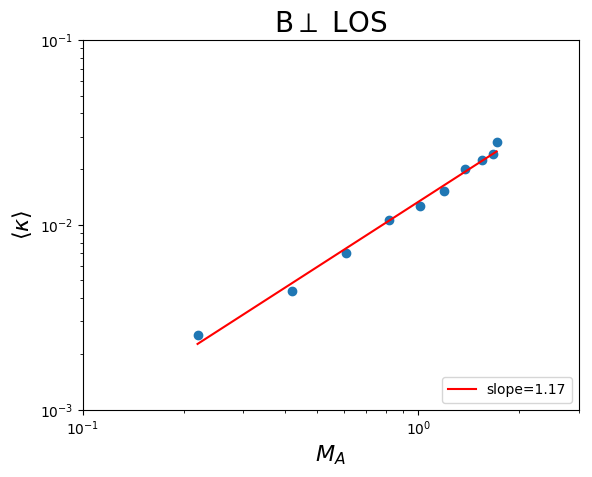}
\includegraphics[width=0.48\linewidth]{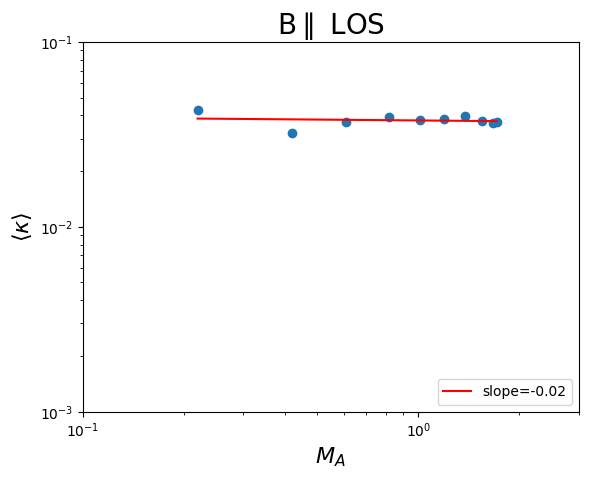}
\includegraphics[width=0.48\linewidth]{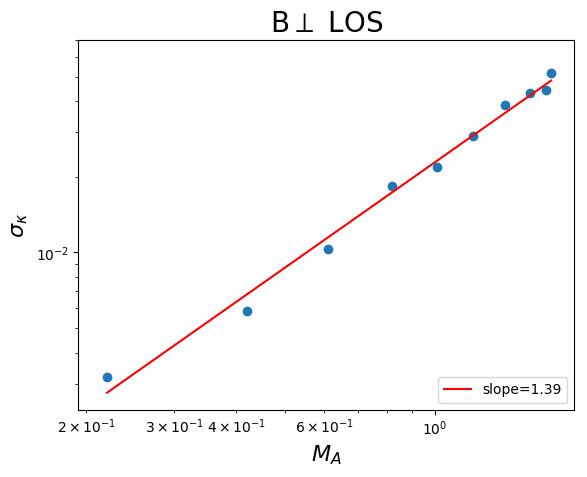}
\includegraphics[width=0.48\linewidth]{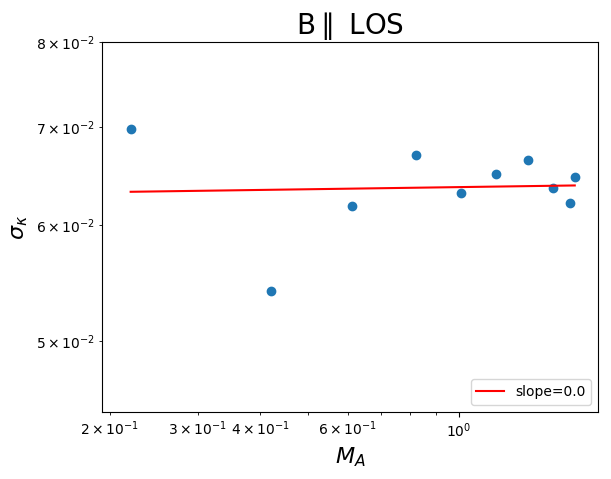}
\includegraphics[width=0.48\linewidth]{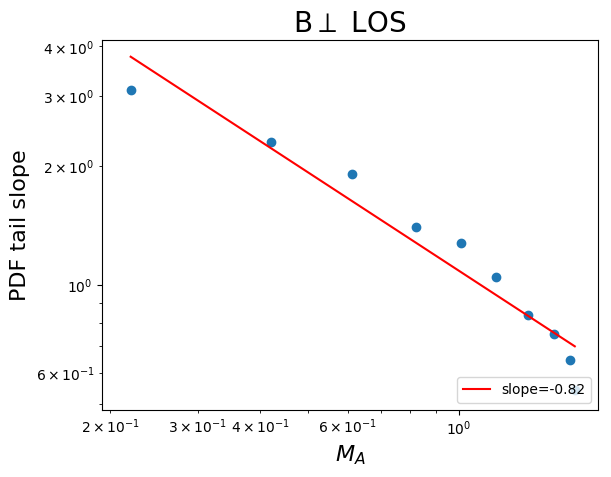}
\includegraphics[width=0.48\linewidth]{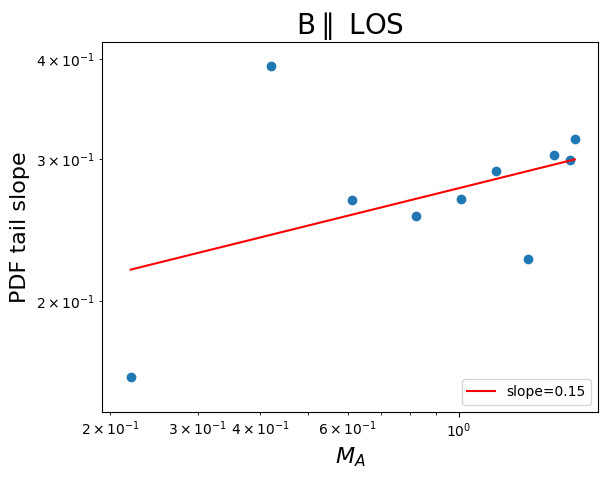}
\caption{\label{fig:curvature_2d} Six panels of figures showing the plot of $\langle \kappa \rangle$ (upper row),  $\sigma_\kappa$ (middle row) and (absolute value of) the slope of the PDF tail (lower row) as functions of $M_A$ when we project the numerical cube perpendicular to the magnetic field (left column) and parallel to the magnetic field (right column) respectively. We provide the slope of the fitting line for each figure.}
\end{figure*}

We would like to see whether the dependencies of \label{subsec:3d} would be seen in observation when we view the cloud parallel or perpendicular to the line of sight. Figure \ref{fig:curvature_2d} shows how $\langle \kappa \rangle$, $\sigma_\kappa$ and the slope of the PDF tail would respond as a function of $M_A$ when we project the Stokes parameter perpendicular and parallel to the mean magnetic field using the procedure laid in \S \ref{sec:method}.

We immediately recognize there are different behavior for the relation of $\langle \kappa \rangle$ to $M_A$ in 2D and 3D. In 3D we have a two-section power law that has a cut-off of $M_A=1$. However, such a power-law is not seen in both cases of $B\perp$ line of sight (LOS) and $B\parallel$ LOS. In fact, in the case of $B\perp$ LOS we see that a uniform power-law describes the data better, which has $\langle \kappa \rangle \propto M_A^{1.17}$. Nevertheless it is shallower than the section of the power-law that we have in $M_A<1$ in \S \ref{subsec:3d} yet steeper than the other section of the power-law that we have in $M_A>1$. A very similar effect also happens to $\sigma_\kappa$ that, only a single power-law would be sufficient to describe the data here ($\sigma_\kappa \propto M_A^{1.39}$).

We also see that the previously seen power law disappears when we project the magnetic field data along its mean field direction, i.e. $B\parallel$ LOS. In this scenario both $\langle \kappa \rangle$ and $\sigma_\kappa$ are more or less a constant of $M_A$. This is expected since we should only see the hydrodynamic nature of the magnetized turbulence if we are observing it along its mean field direction. We see that the value of $\langle \kappa \rangle$ arrives at the maximum value allowed in the curvature algorithm (See \S \ref{subsec:lagrangian}). This suggests that the measured Alfvenic Mach number is actually the perpendicular Mach number $M_{A,perp}=M_A\cos\theta$, where $\theta$ is the angle between the mean magnetic field and the line of sight. 

In the case of the power-law tail slope of the PDF, we see that there still exists a power law between the PDF slope and $M_A$. One interesting thing to note here is, the values of the power-law tail slope actually becomes more disperse when the synthetic maps are obtained by projected perpendicular to the mean magnetic field directions than those in the 3D. When we view the numerical cube with its mean field parallel to the line of sight, we also see the PDF tail slope has different responses as a function of $M_A$ compared to that when $B\perp$ LOS. By using these tools together, it is possible to extract the Alfvenic Mach number in the range of $M_A\in(0.2,1.7)$.

\section{Relation of gradient curvature, magnetic field line curvature and magnetization}
\label{sec:vgt}

Aside from the curvature of magnetic field lines, we could also apply the curvature technique to method that could potentially trace magnetic field lines. The Velocity Gradient technique \citep{YL17a,YL17b,LY18a} is a very powerful technique in probing magnetic field using gradient statistics of turbulence observables. Theoretically for a local magnetic field ${\bf B}$ we expect the term $B\cdot \nabla v$ to be statistically zero as long as the area of sampling is large enough, which is shown in the series of VGT literature. Under the framework of curvature, it is expected to see that the curvature of velocity gradients would be statistically comparable to that of magnetic field curvature. Since we are down-sampling the data with the sub-block averaging \citep{YL17a}, the PDF would not have enough sample to plot even with the smallest block size allowed. As a result we spare the discussion on the PDF tail slope in the sections that involve the block averaging. Here we select a block size of $72$ pixels as case study of how good gradient curvature could represent magnetization

\begin{figure*}[th]
\centering
\includegraphics[width=0.48\linewidth]{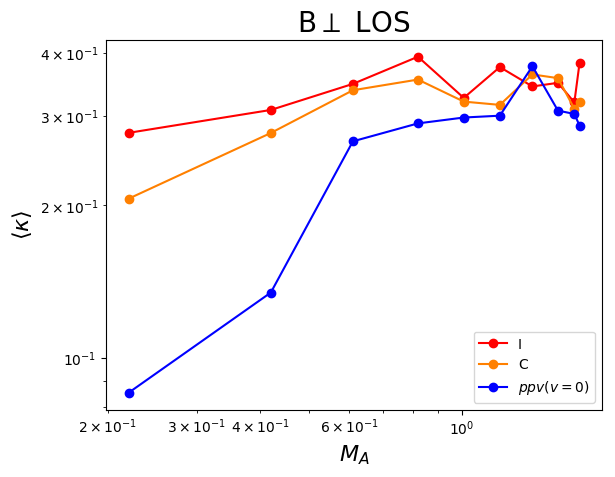}
\includegraphics[width=0.48\linewidth]{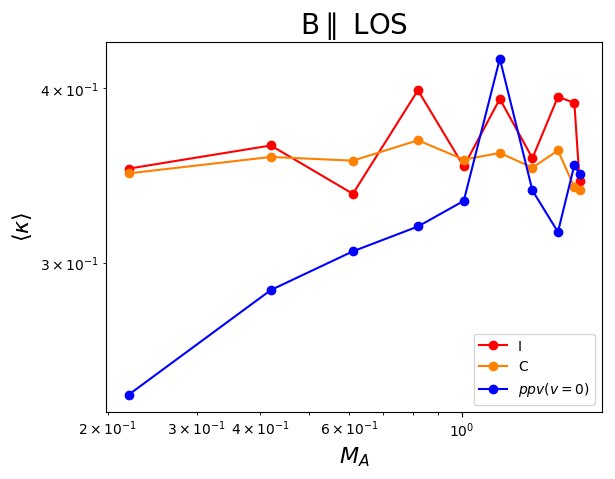}
\includegraphics[width=0.48\linewidth]{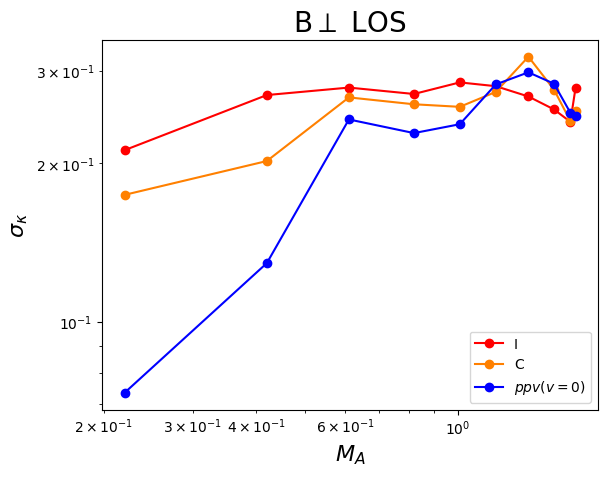}
\includegraphics[width=0.48\linewidth]{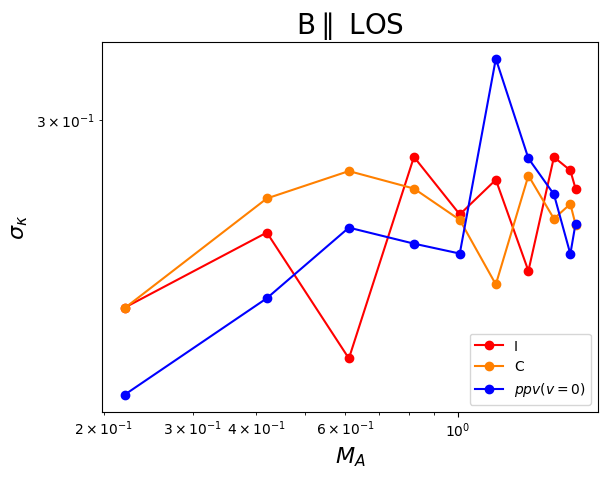}
\caption{\label{fig:curvature_ma_vgt} Six figures showing how the curvature of intensity gradients (red), velocity centroid gradients (orange) and velocity channel with $v=0$ (blue) are related to $M_S$ under three statistical parameters $\langle \kappa\rangle$ (left column) and $\sigma_\kappa$ (right column) for both $B\perp$ LOS (upper row) and $B\parallel$ LOS (lower row). See \S \ref{subsec:syn} for the definition of the observables. }
\end{figure*}

We show the curvature of magnetic field as probed by the Velocity Gradient Technique in Fig. \ref{fig:curvature_ma_vgt}.  Notice that the gradient method in the recipe of \cite{YL17a} would introduce natural dispersion which has been discussed also in \cite{LYH18}. In the concept of curvature that means the minimum curvature attained by gradients of these observables are not zero, which has been reflected as the "base" of the top-base method in \cite{LYH18}. With enough resolution in simulation and statistically sufficient block-sampling this minimum curvature would eventually go to zero.

While there are different dependencies of the statistical measures, there are few interesting properties in Fig. \ref{fig:curvature_ma_vgt} that are consistent to previous sections or works on VGT. For instance, there is a generally growing trend for both  $\langle \kappa\rangle$ and $\sigma_\kappa$ when $M_A$ increases, which suggests that the gradients of these observables are indeed correlated to the magnetic field curvature. Counting the factor of non-zero minimum curvature for gradients, it is possible to correlate the curvature of gradients of observables to the curvature of magnetic field. Second, it is very apparent that the curvature of intensity gradients are significantly larger than that of the velocity centroid and channel gradients especially in the case of $M_A<1$. This is consistent to the previous VGT finding that centroids are generally better than intensity gradients \citep{YL17a,YL17b}, channels are even better in representing the magnetic field structure compared to centroids and intensity gradients \cite{LYH18,PCA} , and also intensity gradients are suffered by shocks and self-gravity \citep{YL17b,IGVHRO}. Third, there is a significant flattening of both $\langle \kappa\rangle$ and $\sigma_\kappa$ of all three observables as $M_A\ge 1$. This is because the turbulent scaling are different in the case of $M_A<1$ and $M_A>1$ (See \citealt{Lazarian2006} for a summary of them).

\section{Contribution of different modes towards curvature of gradients}
\label{subsec:modes}

The concept of MHD modes are crucial in understanding the geometry of magnetic field lines in interstellar media. In compressible turbulence it is found numerically that Alfven modes dominate. In the case of 3D MHD compressible turbulence with the mean field being a straight line, the Alfven mode is believed to be divergence free (\citealt{CL02,CL03,Lazarian2006}, See also \S \ref{subsec:diva} and \S \ref{subsec:modenew} for an alternative discussion). The two compressible modes could then be represented by linear combinations of the so-called parallel (to the local magnetic field, see \S \ref{sec:theory}) and perpendicular components of wave vector. Under this assumption, \cite{CL03} developed a technique in obtaining the unit vectors of the three MHD modes and tested the expected dependence theoretically derived in \cite{GS95} that $k_\parallel \propto k_\perp^{2/3}$. One of the very important discovery from \cite{CL03} is that there is little interference between the incompressible Alfven modes and the compressible slow and fast modes during the whole simulations, suggesting the coupling between Alfven and other modes are weak. The incompressibiltiy of Alfven mode also suggests that the only way to store energy into Alfven mode is to bend it, i.e. introduce curvature to it. While it is also possible to bend the other two modes, the compressibility of these modes act as a spring that can absorb energy when bending them. Therefore it is obvious that Alfven mode would store much more energy due to curvature than the other two modes. Combining with the facts that Alfven modes dominate over the two other modes, we expect that the curvature would be mostly contributed by the Alfven modes. In below we shall examine the velocity gradient curvature induced by different modes as a function of Alfvenic Mach number.

The mode decomposition method for {\it single fluid polytropic MHD turbulence} is described in \cite{CL02} and further elaborated in \cite{CL03}. The three MHD modes, namely the Alfven, slow and fast modes can be decomposed in the Fourier space with respect to the local mean magnetic field ${\bf B_0}$ by:
\begin{equation}
    \begin{aligned}
    {\bf k} &= k_\parallel \hat{\bf B}_0 + k_\perp \hat{\bf k}_\perp = k_\parallel \hat{\bf k}_\parallel + k_\perp \hat{\bf k}_\perp\\
    \hat{\bf \zeta}_A &= \hat{\bf k}_\perp \times \hat{\bf k}_\parallel\\
    \hat{\bf \zeta}_S &\propto (-1+\alpha-\sqrt{D})k_\parallel \hat{\bf k}_\parallel + (1+\alpha-\sqrt{D})k_\perp \hat{\bf k}_\perp\\
    \hat{\bf \zeta}_F &\propto (-1+\alpha+\sqrt{D})k_\parallel \hat{\bf k}_\parallel + (1+\alpha+\sqrt{D})k_\perp \hat{\bf k}_\perp
    \end{aligned}
\end{equation}
These unit vectors are mutually orthogonal and spans over $\mathscr{R}^3$. The decomposition is valid when the physical scale is larger than the 1st decoupling scale $k_{dec,ni}$ as neutrals and ions can be considered to be one co-moving species instead of two (See Xu et.al 2015,2016). The decomposition starts to become physically unjustified as between the two decoupling scales $k_{dec,ni}=\gamma_d \rho_i<\gamma_d\rho_n=k_{dec,in}$ there are three more damping modes for the two velocities ${\bf v}_{i,n}$ that characterize the two-fluid partially coupling MHD equations. In our case we are mostly in the diffuse ISM regime, therefore we could stick with the single fluid mode decomposition for the rest of the current section.

\begin{figure*}[th]
\centering
\includegraphics[width=0.48\linewidth]{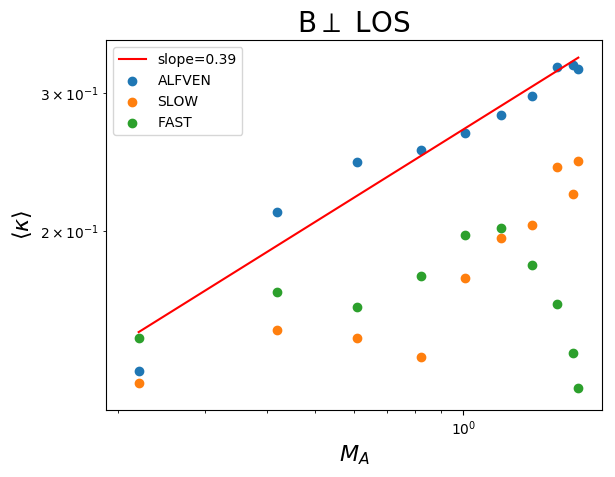}
\includegraphics[width=0.48\linewidth]{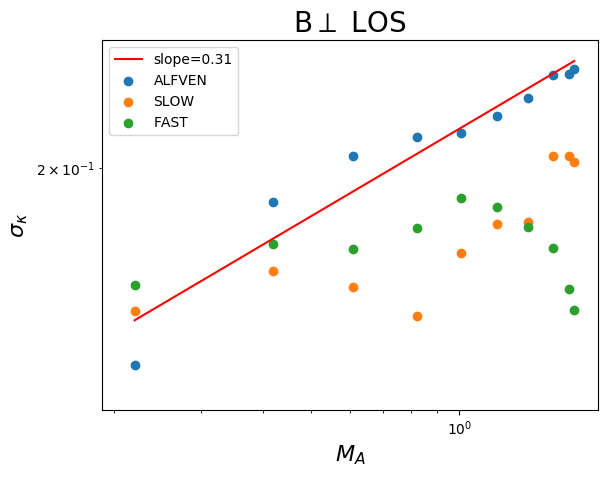}
\caption{\label{fig:curvature_ma_fsa} Two figures showing how $\langle \kappa \rangle$ and $\sigma_\kappa$ of the gradients of Alfven (blue), slow (orange) and fast (green) modes behave as a function of $M_A$. We also provide the fitting line for the Alfven mode which is a more apparent power-law relation in these two figures.}
\end{figure*}

The line of sight component of the three modes are projected perpendicularly so that we would not be interfered by the line of sight effect (See \S \ref{sec:stats}). The projection is effectively computing the velocity centroid of these three modes assuming the density as a constant. The assumption is necessarily since the density fluctuation corresponding to Alfven mode is zero but this is not the case for slow and the fast mode. For a fair comparison we only consider their velocity fluctuations and see how would the three modes behave. In this scenario we find that a block size of 18 pixels would fulfill the block-averaging condition \citep{YL17a}.

Fig. \ref{fig:curvature_ma_fsa} shows how $\langle \kappa \rangle$ and $\sigma_\kappa$ of the gradients of Alfven (blue), slow (orange) and fast (green) modes behave as a function of $M_A$. We also provide the fitting line for the Alfven mode which is a more apparent power-law relation in these two figures. We can immediately see a few things from these figures. First, the result from Fig. \ref{fig:curvature_ma_fsa} shows Alfven mode indeed has a higher curvature value compared to the other two modes, with the only exception of $M_A=0.2$ case (huge-0). The reason of the latter is because in the case of very low $\beta$ the compression perpendicular to the line of sight is simply much stronger. As a result one could see dominance of fast modes which would be almost $\parallel k_{\perp}$ in the case of $M_A=0.2$ since the respective $\beta$ value is $\ll1$ (See Table \ref{tab:sim} and also Appendix of \citealt{CL03}). Another thing that we see from Fig. \ref{fig:curvature_ma_fsa} is that, while there is no apparent power-law fit that could be found between both $\langle \kappa \rangle$ and $\sigma_\kappa$ to $M_A$ for the two compressible modes, there exists a nicely fit power law for Alfven mode to $M_A$ but with a much flatter slope, contrary to the absence of power law in what we see from \S \ref{sec:vgt}. The filtering of slow and fast modes are well known to improve hunting of anisotropy \citep{2010ApJ...720..742K} and also performance of the velocity gradient technique \cite{Letal17,LYH18}. The existence of the power law on Alfven modes to $M_A$ would indicate the magnetization could be estimated if the compressible modes are filtered.

\section{Application to observational data}
\label{sec:obs}
We test our method in \S \ref{sec:stats} in the recently available Planck data \citep{2018arXiv180706205P}. We use the 353 GHz full-sky map (R3.01) and obtain the polarization angle $\phi=0.5atan_2(U/Q)$ from the data. We then compute the circular dispersion of polarization angles $\delta \phi$ and the curvature quantities $\langle \kappa \rangle$ and $\sigma_\kappa$ as proposed in \S \ref{sec:stats}. They are all computed in a block-size of (4.17 degrees)$^2$. Since these three parameters are all increase with $M_A$, we could see whether the approach employing curvature is useful. We do this by comparing the spatial correlation of the curvature statistics to that of the dispersion of angles.

\begin{figure}[th]
\centering
\includegraphics[width=0.97\linewidth]{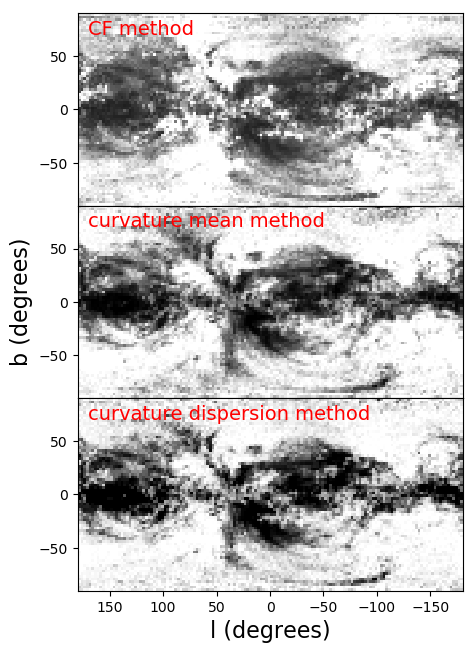}
\caption{\label{fig:planck_1} Three panels of figures showing the dispersion of angles $\delta \phi$ (\citealt{CF53}, top panel), the mean $\langle \kappa \rangle$ and dispersion of curvature $\sigma_\kappa$  method (middle and lower panel\S \ref{sec:stats}) with a block size of (4.17 degrees)$^2$. The color scale is set such that a lighter color indicates a larger value of that quantity and scales up within one standard deviation. }
\end{figure}

Fig. \ref{fig:planck_1} shows how these parameter are distributed in the full sky. One could see visually that these three methods are spatially correlated. In the following we employ the normalized cross-correlation function (NCC) $NCC(a,b) = Cov(a,b)/\sigma_a \sigma_b$, where $Cov$ is the covariance function, as a tool to compare the similarity of the two two-dimensional maps. The NCC has a range of $[-1,1]$, with $NCC>0$ meaning the two maps have positive correlation and vice versa. We compute the NCC function between the dispersion of angles to both mean $\langle \kappa \rangle$ and dispersion of curvature $\sigma_\kappa$ and see that:
\begin{equation}
    \begin{aligned}
    NCC(\delta \phi, \langle \kappa \rangle) &\sim 0.63\\
    NCC(\delta \phi, \sigma_\kappa) &\sim 0.76\\
    \end{aligned}
\end{equation}

\begin{figure*}[th]
\centering
\includegraphics[width=0.97\linewidth]{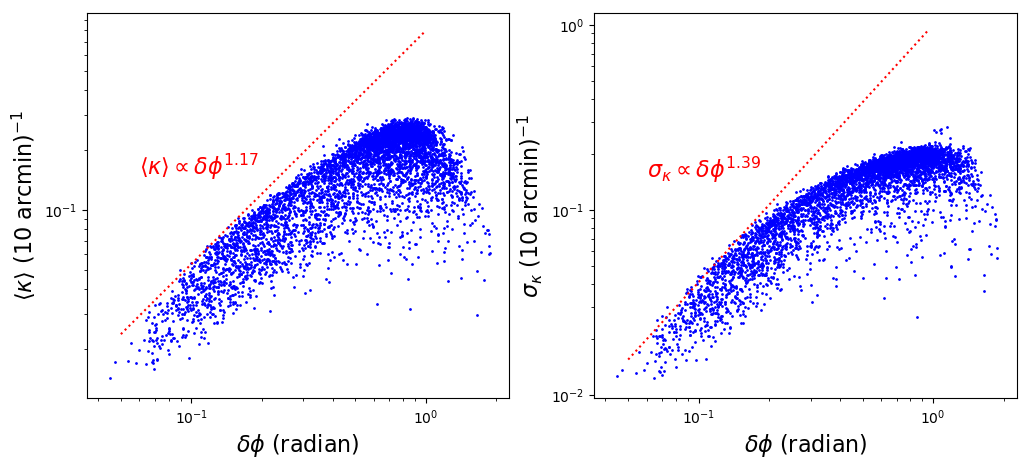}
\caption{\label{fig:planck_2} Two figures showing the scatter plot of $\langle \kappa \rangle$-$\delta\phi$ (left) and $\sigma_\kappa $-$\delta\phi$ (right) using the 353 Ghz Planck Polarization data at a block size of (4.17 degrees)$^2$. The two trend lines are added according to the prediction in \S \ref{subsec:stokes}.}.
\end{figure*}

We would also compare the two measurements locally. We know that $\delta \phi \propto M_A$ and from \S \ref{subsec:stokes} we know that $\langle \kappa \rangle\propto M_A^{1.17} $ and $\sigma_\kappa \propto M_A^{1.39}$. Therefore we expect $\langle \kappa \rangle\propto \delta \phi^{1.17} $ and $\sigma_\kappa\propto \delta \phi^{1.39}$. Fig. \ref{fig:planck_2} shows how the scatter plot of $\langle \kappa \rangle$ (left) and $\sigma_\kappa$ (right) behave with respect to $\delta\phi$. We added the trends as predicted in \S \ref{subsec:stokes} and see that the trend lines in both figures predict the shape of both scatter plot when $\delta \phi<0.2$, which is consistent the numerical exploration in \S \ref{subsec:stokes}. For $\delta \phi>0.2 \text{radian} \sim 11.4^o$, both figures have a significant deviation from the curvature statistics. However generally both $\langle \kappa \rangle$ and $\sigma_\kappa$ have a positive relation to $\delta \phi$. This shows that the statistical measure of $\kappa$ is a valid measurement of magnetization.

\section{Discussion}
\label{sec:discussions}
\subsection{The use of torsion in studying magnetic field}
\label{subsec:torsion}

Aside from curvature, torsion is also a very important geometric quantity for the magnetic field. While the use of torsion is less popular in literature,  we should not underestimate the importance of torsion since it records how sharply the magnetic field lines twist out of the plane of curvature. The concept of torsion is especially useful when we are dealing with the natural oscillation modes in magnetized turbulence, namely the fast, slow and Alfvenic modes (See \citealt{2003matu.book.....B,CL03}). In fact, when a magnetic field mode is propagating along some directions, the speed of rotation of the eddy that is {\it along the azimuthal direction} of the propagating directions records the torsion of the mode. The signature of magnetic field rotates when propagating is especially important when we are studying modes in MHD turbulence. 

Curvature and torsion could be easily visualized when the magnetic field lines have some special geometry. For instance, if the magnetic field line could be parametrized by a circular helix $L_B(t) = \left( a\cos(t), a\sin(t), bt\right)$, then the curvature and torsion of this magnetic field line are simply $\kappa=a/\sqrt{a^2+b^2}$ and $\tau=b/\sqrt{a^2+b^2}$, respectively. From Eq. \ref{eq:FSF} we could have a formula for the signed torsion:
\begin{equation}
    \tau   = \hat{b}\cdot\frac{d\hat{n}}{ds}
    \label{eq:CT2}
\end{equation}

The estimation of curvature and torsion would be also very important in advancing the velocity gradient technique since essentially Eq.\ref{eq:FSF} describes the behavior of first three derivatives of the orientation of magnetic fields. In the Velocity Gradient Technique we approximated $\bf \hat{t}$ by block averaging \citep{YL17a}. The curvature could be possibly approximated by the magnetization technique by \citep{LYH18} with Fig. \ref{fig:curvature_ma_3D} in 3D or Fig. \ref{fig:curvature_2d} in 2D. Numerically if we know both the tangent and curvature, we could then solve the magnetic field line equation by Eq.\ref{eq:FSF} and Eq.\ref{eq:integrator}. By studying the properties of torsion field, we would know how the second derivative of the orientation of magnetic field would look like and that would provide us a geometrically more plausible way in reconstructing the magnetic field directions.

\subsection{Divergence of Alfven mode}
\label{subsec:diva}

One of the very important applications of the Frenet-Serret frame derived from magnetic field line is to recognize how the three MHD modes are related to the curvature and torsion of magnetic lines. In below, we shall discuss how the three MHD modes would be related to the local Frenet-Serret frame of magnetic field by the theory of MHD turbulence \citep{GS95,LV99}. In the case of \cite{CL03} decomposition the mean field has to be locally a straight line\footnote{\torefereeone The "locality" issue in \cite{CL03} is hard in implement in numerical simulations. In theory, if one could sample a small enough space in the turbulence cube provided that the space has enough resolution for the Fourier transform, the locality could be obtained within that small enough space, which is the essence of the \cite{CL03} decomposition. However, due to the restrictions of resolutions, the statistics of a very small region in numerical simulations are dominated by dissipations. \cite{2010ApJ...720..742K} later uses the method of wavelets to localize the \cite{CL03} formalism.}. In this scenario there is an ambiguity in defining the directions of both $\hat{n}$ and $\hat{b}$. One could simply assign the Alfven wave unit vector to be along one of these directions. However in the case of a mean magnetic field with non-zero curvature, the effect of frame changes in the local regions has to be taken into account and we have to drop some assumptions that were valid in usual compressible 3D magnetized turbulence. For instance, the assumption that Alfven mode to be divergence free is not correct when the local mean field has non-zero curvature. \cite{1985P&SS...33..127S} proposes that the Alfven mode has a non-zero divergence related to the curvature of the mean magnetic field:
\begin{equation}
    \nabla \cdot \hat{\zeta}_A \sim \kappa\hat{\zeta}_A\cdot \hat{n}
    \label{eq:divza}
\end{equation}
where $\hat{\zeta}_A$ is the Alfven mode unit displacement vector. Following \cite{1985P&SS...33..127S}, there is a source term for the slow wave equation to be proportional to $\kappa\hat{\zeta}_A\cdot \hat{n}$, while the Alfven mode wave equation has similar term that is $\propto \kappa \hat{n}$. Physically the bent magnetic field would induce a change of fluid element volume towards the center of curvature, which its size is proportional to $\kappa$. As a result, the plasma pressure due to the compression of volume is changed under this magnetic field geometry, which would drive the slow mode to oscillate along with the azimuthal direction of Alfven mode. However the Alfven mode in this situation is no longer solendoial (See Eq. 23 or Fig 3 of \citealt{1985P&SS...33..127S}), which is different from the case when we perform mode decomposition with a mean field with $\kappa=0$ (See \citealt{CL03} for a discussion of MHD modes). Unlike \cite{CL03} where the mean magnetic field was approximated by a straight line, the study in \cite{1985P&SS...33..127S} points out that in the presence of uniform non-zero curvature magnetic field, the transfer of energy between Alfven and slow modes increases. This effect, is however reduced by the Alfvenic cascade happening in one eddy turnover time, which makes the effect important only when the curvature is comparable with the wave number of the perturbations under consideration. 

\subsection{The picture of MHD modes under strongly curved magnetic field}
\label{subsec:modenew}
The expression Eq.\ref{eq:divza} suggests that the Alfven mode wavevector, which lies on the plane spanned by the normal and the binormal vector, would rotate as a function of curvature with its rotation axis aligned with the tangent vector ${\bf t}={\bf B}_0$. Eq.\ref{eq:divza} suggests that if $\lambda_{\perp}$ is the perpendicular wavelength and $\phi_k=\cos^{-1}(\hat{k}\cdot \hat{b})$, then the angle of rotation $\phi_A$ of the Alfven mode with respect to the tangent vector is given by:
\begin{equation}
    \cos\phi_A \sim \frac{\cos\phi_k}{\lambda_\perp \kappa}
\end{equation}
Therefore for a magnetized turbulent system with a non-zero mean magnetic field curvature, one could (1) compute the components of wavevector with respect to the Fernet-Serret frame (2) compute the orientation of MHD modes according to \cite{CL03} and represent these eigenvectors by Fernet-Serret frame (3) rotate the three MHD mode vectors with the rotation matrix $R(\hat{t},\phi_A)$ where $\hat{t}$ is the rotation axis. This method allows one to use the formulation of \cite{CL03} to compute the MHD modes.

\subsection{Importance of gravity to magnetic field}
\label{subsec:gravity}

The important insight of increasing coherent coupling between Alfven and slow modes in the presence of non-zero curvature mean field suggests that using the concept of modes in systems that has non-zero mean magnetic field curvature should be taken in caution. In \S \ref{subsec:math}, we see that the curvature naturally introduce a length scale related to the radius of curvature $r_c = \kappa^{-1}$. Above the aforementioned length scale Alfven and slow modes are coupled while below that they act independently. In the case that there are no external forces holding the magnetic field, the magnetic field curvature would restore to infinity so that the magnetic field will become a straight line when the magnetic energy from the curvature is all transferred to the dissipation of Alfven and slow modes. 

However, in real astrophysical scenarios, there are external forces that can keep magnetic fields bent. One of the best examples would be self-gravity. It is shown observationally by \cite{2015Natur.520..518L} that the curvature of magnetic field could be used in estimating the magnetic field strength on a self-similar molecular cloud, suggesting that the curvature of magnetic field does not dissipate to Alfven and slow mode driving in a self-gravitating system. In fact, it is easy to imagine that the gravitational field provides support for the curvature of magnetic field. Following \cite{2015Natur.520..518L} , for a spherical self-gravitating object of constant density $\rho$, radius $R$ and mean magnetic field strength $B$, along the normal direction of magnetic field:
\begin{equation}
    \frac{B^2}{R} \sim \frac{4\pi}{3}G\rho^2R
    \label{eq:curvature_condition}
\end{equation}
which one would have $\kappa (v_A t_{ff})\sim 1$, where $t_{ff} \sim (G\rho)^{-1/2}$ is the free fall time. When $r_c = \kappa^{-1} \gg v_A t_{ff}$ then magnetic field dominates over gravity, vice versa. This suggests that curvature could be supported by gravity if Eq.\ref{eq:curvature_condition} is satisfied.

Under the scenario that $\kappa (v_A t_{ff})\sim 1$, the gravitational energy would become the primary energy source in driving both (non-incompressible) Alfven and slow waves. When the density of the self-gravitating object increases, it is expected to have a run-off effect by having a larger curvature and a much stronger coupling between the Alfven and slow modes. Moreover, the dependence of curvature to magnetic field changes from $\kappa \propto B^{-2}$ in non-gravitating systems to $\kappa \propto \rho B^{-1}$ in self-gravitating systems.

\section{Conclusions}
\label{sec:conclusion}
The use of geometrical properties of magnetic field lines and interaction with magnetized turbulence would help us advancing both the theory of MHD turbulence and also introducing ways in studying magnetic fields in observation. In this work, we explore the statistical properties of magnetic field line curvature $\kappa$ in compressible magnetized turbulence. To summarize:
\begin{enumerate}
    \item We propose an algorithm in computing the curvature of both gradients of a scalar field and a vector field (\S \ref{subsec:lagrangian})
    \item We study the mean value and the standard deviation of magnetic field line curvature and identify the power law relation of the two quantities with the magnetization. (\S \ref{subsec:3d}).
    \item We also obtain the power law relation of the spectral index of the histogram of curvature with the Alfvenic Mach number (\S \ref{subsec:3d})
    \item The power-laws can also be seen in observation with the exception of degenerate cases when mean magnetic field is either parallel or perpendicular to the line of sight (\S \ref{subsec:stokes})
    \item The curvature method can be used in advancing the Velocity Gradient Technique and predicts the magnetization based on the gradients of observables (\S \ref{sec:vgt})
    \item The MHD mode analysis shows that Alfven mode is the dominant curvature contribution towards the magnetic field lines traced by velocity gradients (\S \ref{subsec:modes})
    \item We test our prediction of power law in \S \ref{subsec:stokes} in observation and see the same relation of $\kappa$ to $\delta \phi$ when the latter is small (\S \ref{sec:obs}).
    \item We discuss the modifications of the physics of Alfven modes when the background magnetic field has {\torefereeone significant} curvature.  (\S \ref{subsec:diva}). 
\end{enumerate}

{\bf Acknowledgment.} K.H.Y acknowledge Korea Astronomy and Space Science Institute on November 2018 for hospitality in inspiring the curvature algorithm. We acknowledge the support the NSF AST 1816234 and NASA TCAN 144AAG1967.  AL thanks the Center for Computational Astrophysics (CCA) for its hospitality. This research used resources of both Center for High Throughput Computing (CHTC) at the University of Wisconsin and National Energy Research Scientific Computing Center (NERSC), a U.S. Department of Energy Office of Science User Facility operated under Contract No. DE-AC02-05CH11231, as allocated by TCAN 144AAG1967. 

\appendix
\section{\torefereeone The convergence test for the curvature algorithm}
\label{appendix:convergence}
\begin{figure*}[th]
\centering
\includegraphics[width=0.97\linewidth]{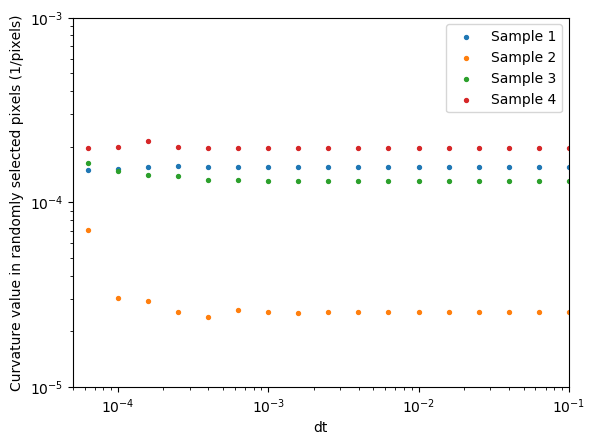}
\caption{\label{fig:convergence_test}\torefereeone A figure showing how the curvature value for 4 randomly drawn pixels in the simulation huge-0 depends on the step size $dt$.}.
\end{figure*}

{\torefereeone In this section we perform a convergence test and see if the algorithm that we developed in \S \ref{subsec:lagrangian} would converge as we change the only parameter $dt$ in the algorithm. We test a wide range of the step size and Fig .\ref{fig:convergence_test} shows the value of curvature in 4 randomly drawn pixels in our simulations as a function of step size. We can see that the estimated curvature value is basically constant within the range of the $dt$ that we selected. We therefore believe that our method of obtaining curvature is robust when an appropriate $dt$ is selected. }

\end{document}